\documentclass[aps,prb,twocolumn,superscriptaddress,a4paper]{revtex4-1}
\usepackage{color}
\usepackage{graphicx}
\usepackage{latexsym}
\usepackage{dcolumn}
\usepackage{amsmath}
\usepackage{amssymb}
\usepackage[normalem]{ulem}

\newcommand{\be}{\begin{equation}}
\newcommand{\ee}{\end{equation}}
\newcommand{\bea}{\begin{eqnarray}}
\newcommand{\eea}{\end{eqnarray}}

\def\a{\alpha}
\def\b{\beta}
\def\g{\gamma}

\def\d{\delta}
\def\D{\Delta}
\def\e{\epsilon}
\def\ve{\varepsilon}

\def\p{\pi}

\def\w{\omega}

\def\callZ{\mbox{$\mathcal{Z}$}}

\def\1op{\hat{\mathbbm{1}}}
\def\nn{\nonumber}
\begin{document}

\title{Time-dependent i-DFT exchange-correlation potentials with 
memory: Applications to the out-of-equilibrium Anderson model}
\author{Stefan Kurth}

\affiliation{Nano-Bio Spectroscopy Group and European Theoretical Spectroscopy 
Facility (ETSF), Dpto. de F\'{i}sica de Materiales,
Universidad del Pa\'{i}s Vasco UPV/EHU, Av. Tolosa 72, 
E-20018 San Sebasti\'{a}n, Spain}
\affiliation{IKERBASQUE, Basque Foundation for Science, Maria Diaz de Haro 3, 
E-48013 Bilbao, Spain}
\affiliation{Donostia International Physics Center (DIPC), Paseo Manuel de
Lardizabal 4, E-20018 San Sebasti\'{a}n, Spain}
\author{Gianluca Stefanucci}
\affiliation{Dipartimento di Fisica, Universit\`{a} di Roma Tor Vergata,
  Via della Ricerca Scientifica 1, 00133 Rome, Italy}
\affiliation{INFN, Sezione di Roma Tor Vergata, Via della Ricerca Scientifica 1, 00133 Roma, Italy}

\date{today}

\begin{abstract}
We have recently put forward a steady-state density 
functional theory (i-DFT) to calculate the transport coefficients 
of quantum junctions. Within i-DFT it is 
possible to obtain the steady density on and the steady current 
through an interacting junction
using a fictitious noninteracting junction subject to an 
effective gate and bias potential. In this work we extend i-DFT to 
the time domain for the single-impurity Anderson model. By a reverse 
engineering procedure we extract the exchange-correlation (xc)
potential and xc bias at temperatures above the Kondo temperature 
$T_{\rm K}$. 
The derivation is based on a generalization of a recent paper by
Dittmann et al. [arXiv:1706.04547]. Interestingly the time-dependent (TD)  
i-DFT potentials depend on the system's history only through the first 
time-derivative of the density. We perform numerical simulations of the early 
transient current and investigate  the role of the 
history dependence. We also empirically extend the history-dependent TD 
i-DFT potentials to temperatures below $T_{\rm K}$. For this purpose 
we use a recently 
proposed parametrization of the  i-DFT 
potentials which yields highly accurate results in 
the steady state.
\end{abstract}
\maketitle

\section{Introduction}

The enormous success of density functional theory (DFT)
\cite{HohenbergKohn:64,KohnSham:65,DreizlerGross:90} in describing equilibrium
properties of weakly correlated systems has triggered interest towards the
development of DFT approximations to deal with strong electronic correlations
\cite{LimaOliveiraCapelle:02,LimaSilvaOliveiraCapelle:03,MoriSanchezCohenYang:09,XianlongChenTokatlyKurth:12,CapelleCampo:13,MaletGoriGiorgi:12,MirtschinkSeidlGoriGiorgi:13,BroscoYingLorenzana:13,YingBroscoLorenzana:14,CarrascalFerrerSmithBurke:15}.
The extension of DFT to non-equilibrium situations has been put forward 
by Runge and Gross in a milestone paper from 1984~\cite{RungeGross:84}.
Time-dependent DFT (TDDFT) has been successfully
applied to atoms, molecu\-les and solids \cite{Ullrich:12,Maitra:16} and, more
recently, to strongly correlated systems like, e.g., Hubbard wires or
nanoclusters \cite{Verdozzi:08,KKPV.2013,FFTAKR:13,FuksMaitra:14,FuksMaitra:14-2,CarrascalFerrerMaitraBurke:18} using 
approximations borrowed from static DFT.
These approximations have also been extended to 
nanoscale systems in contact with metallic leads to investigate transient
currents and more generally transport coefficients \cite{kskvg.2010,UimonenKhosraviStanStefanucciKurthLeeuwenGross:11,KhosraviUimonenStanStefanucciKurthLeeuwenGross:12,sds.2013,KarlssonVerdozzi:16,YPKSD.2016}. 

The single impurity Anderson model (SIAM) 
is the simplest model exhibiting nontrivial 
strong correlations effects, namely the formation of the Kondo singlet at temperatures 
below the Kondo temperature $T_{\rm K}$ and the Cou\-lomb blockade (CB)
phenomenon at temperatures lower than the charging energy (or on-site 
repulsion) $U$. 
Both effects leave clear fingerprints on the steady-state and 
time-dependent transport properties of the SIAM, and until a few 
years ago there were no suitable DFT approximations 
for their description. 

In 2011 we and two other groups independently proposed a DFT 
exchange-correlation (xc) potential able 
to capture the Kondo plateau in the zero-bias conductance 
at vanishing temperature~\cite{sk.2011,blbs.2012,tse.2012}. In 
Ref.~\cite{sk.2011} the xc potential $v_{\rm xc}$ was obtained by reverse 
engineering the exact solution of the  SIAM with the uncontacted impurity 
(equivalent to 
the single-site Hubbard model). However, it 
was soon realized that such a $v_{\rm xc}$ fails 
dramatically at temperatures $T>T_{\rm K}$ and/or at finite bias, 
the cause of the failure being the lack of {\em dynamical xc 
effects}~\cite{ks.2013,StefanucciKurth-PSSB}. 
In fact, DFT is an {\em equilibrium} theory and it is 
not supposed to describe nonequilibrium properties like transport 
coefficients. The fortunate success in reproducing the Kondo plateau 
is a direct consequence of the Friedel sum rule~\cite{mera-1,mera-2},
according to which the zero-bias 
and zero-temperature conductance is completely determined by 
the ground state density of the impurity~\cite{Friedel2001,Langreth:66}. 

A consistent framework to deal with
quantum transport is TDDFT where the xc potential at a certain space-time point 
depends on the density everywhere and at all previous times
\cite{Ullrich:12,Maitra:16}. This 
memory dependence is what we mean by dynamical xc 
effects~\cite{ks.2013,StefanucciKurth-PSSB,LiuBurkePhysRevB.91.245158}. 
Their inclusion, however, is far 
from trivial since it requires the knowledge of the TD 
density deep inside the leads~\cite{sa-1.2004,sa-2.2004,ewk.2004,ksarg.2005}, 
a portion of the system which is  usually integrated out by an 
embedding procedure. To overcome this complication we formulated
a steady-state density functional theory applicable to any interacting 
junction to extract differential
conductances~\cite{StefanucciKurth:15} as well as equilibrium
spectral functions \cite{JacobKurth:18}. 
In this framework, henceforth named i-DFT,
the basic variables are 
the density on and the steady current through the junction, $(n,I)$, whereas the conjugated 
variables are the potential acting on the junction, or gate voltage 
$v$, 
and the external bias $V$. The map  
$(n,I)\leftrightarrow (v,V)$ is bijective in a finite (gate 
dependent) window around $V=0$ for any finite temperature.

Using the rate equations approach of Ref.~\cite{rate-paper1} we 
obtained the i-DFT xc gate and xc bias of the SIAM by a reverse 
engeeniring procedure and proposed a simple parametrization for 
them~\cite{StefanucciKurth:15}. The rate equations are accurate only 
in the CB regime, or better for temperatures larger than the level 
broadening (hence the Kondo physics is left out).
More recently we have been able to extend the parametrization of the i-DFT 
xc potentials to deal with 
arbitrary temperature $T$ and on-site 
repulsion $U$~\cite{KurthStefanucci:16,KurthStefanucciJPCreview}. 
The results have been shown to agree with those of numerically 
exact techniques, like the functional renormalization group and the 
numerical renormalization gruoup,
in a wide range of temperatures,
on-site repulsions, gate voltages and biases with very high accuracy.

The main merit of the i-DFT formulation  
is the possibility of 
calculating the SIAM differential conductance in {\em any 
regime} at a {\em negligible computational cost}.
Nevertheless, i-DFT is a steady-state theory and time-dependent responses 
are, by construction, left out. The 
extension  of i-DFT to 
the time domain is currently missing. 
For TD i-DFT to be a rigorous framework one should prove that
the TD density $n$ and current $I$ are in a one-to-one correspondence with 
the  TD gate $v$ and voltage $V$. If so, then it would be possible to 
calculate $n$ and $I$ from a 
fictitious noninteracting junction driven by an effective  gate 
$v_{s}=v+v_{\rm Hxc}$ and 
voltage $V_{s}=V+V_{\rm xc}$, where the Hartree-xc (Hxc) gate potential 
$v_{\rm Hxc}=v_{\rm Hartree} + v_{\rm xc}$ and the xc bias $V_{\rm xc}$ 
are universal (i.e., independent of the external potentials) 
functionals of the density and current at all previous 
times. Assuming the existence of the TD i-DFT map, 
 in Ref.~\cite{stefanucci2017ac} we have  calculated the AC differential 
conductance of the SIAM using 
the steady-state i-DFT potentials evaluated at the instantaneous 
current and density (in analogy with the adiabatic 
approximation of TDDFT constructed from 
functionals of ground state DFT).

In this work we show that the TD i-DFT map does exist for the 
SIAM in the Coulomb blockade regime.  We generalize a 
recent paper by Ditt\-mann et al.~\cite{dittmann2017non}
and derive a coupled system of 
equations for the construction of $v_{\rm Hxc}$ and $V_{\rm 
xc}$. We also propose a simple, yet accurate, parametrization 
for these potentials and show that the memory dependence occurs only
through the first time-derivative of the density.

\section{TD i-DFT functional with memory}

\begin{figure}[t]
  \includegraphics[width=0.47\textwidth]{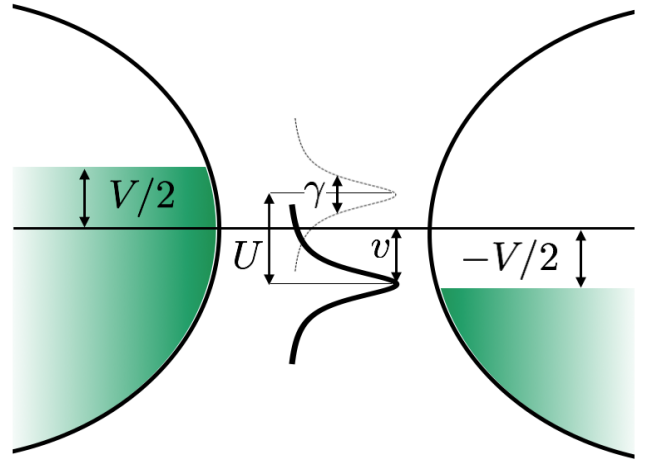}
\caption{Schematic illustration of the SIAM. A left and right lead, 
with bias $V/2$ and $-V/2$, respectively are connected to a single 
level with energy $v$. Due to the coupling to the leads the level is 
broadened by $\g$. $U$ is the energy it costs to add an electron of 
spin up (down) if the impurity is occupied by an electron of spin 
down (up).}
\label{sketch}
\end{figure}

The nonequilibrium SIAM Hamiltonian describes a single impurity level, 
subject to a TD gate $v(t)$,
coupled to a left ($L$) and right ($R$) electronic reservoirs, subject 
to a TD bias $V_{L}(t)=V(t)/2$ and $V_{R}(t)=-V(t)/2$. 
Since the bandwidth of the reservoirs is the largest energy scale,
the density of states in $L$ and $R$ is assumed to be a frequency 
independent constant, i.e., we work in the wide band limit (WBL).
Consequently, the broadening $\g=\g_{L}+\g_{R}$ of the impurity 
level is also frequency independent. The electrons interact
with repulsion energy $U$ only if they are located on the impurity. 
In Fig.~\ref{sketch} we show a schematic illustration of the 
SIAM along with the various energy scales.

We work in the regime $\g\ll T,U$. 
Then, in accordance with Ref.~\cite{rate-paper1} we can write a set 
of coupled equations for the probabilities $P_{0},\,P_{1}$ and $P_{2}$ 
that the impurity is occupied by zero, one (of spin up or down) and two 
electrons respectively:
\bea
\frac{1}{\g}\frac{dP_{0}}{dt}&=&-P_{0}\sum_{\a}f_{\a}(v)+\frac{P_{1}}{2}\sum_{\a}\bar{f}_{\a}(v),
\label{re1}
\\
\frac{1}{\g}\frac{dP_{1}}{dt}&=&-\frac{P_{1}}{2}\sum_{\a}
\left[f_{\a}(v+U)+\bar{f}_{\a}(v)\right]
\\
&&+P_{0}\sum_{\a}f_{\a}(v)+P_{2}\sum_{a}\bar{f}_{\a}(v+U),
\label{re2}
\nn\\
\frac{1}{\g}\frac{dP_{2}}{dt}&=&-P_{2}\sum_{a}\bar{f}_{\a}(v+U)
+\frac{P_{1}}{2}\sum_{\a}f_{\a}(v+U).
\label{re3}
\eea
These are known as rate equations (RE). The sum runs over $\a=L,R$ and 
\be
f_{\a}(\e)\equiv \frac{1}{e^{\b(\e-V_{\a})}+1}
\ee
is the Fermi function of lead $\a$ at inverse temperature $\b=1/T$ and (without 
loss of generality) vanishing chemical potential. The function
$\bar{f}_{\a}(\e)\equiv 1-f_{\a}(\e)$.  The three RE are not linearly 
independent; it is easy to verify that 
\be
P_{0}+P_{1}+P_{2}=1,
\label{norm}
\ee
as it should be. From the probabilities the electronic occupation $N$ 
of the impurity and the current $I_{\a}$ at the $\a$ interface can be 
calculated according to
\be
N=P_{1}+2P_{2},
\label{dens}
\ee
\be
\frac{I_{\a}}{\g}=P_{0}f_{\a}(v)+\frac{P_{1}}{2}
\left[f_{\a}(v+U)-\bar{f}_{\a}(v)\right]-P_{2}\bar{f}_{\a}(v+U).
\label{current}
\ee

The steady-state solution at finite (constant) bias is obtained by setting 
$dP_{i}/dt=0$. Using the first and third RE together with the 
probability normalization Eq.~(\ref{norm}) one finds
\be
P_{0}=\frac{1}{R(0)}\frac{P_{1}}{2}\quad,\quad 
P_{2}=R(U)\frac{P_{1}}{2},
\ee
with
\be
P_{1}=\frac{1}{1+\frac{1}{2R(0)}+\frac{R(U)}{2}}
\ee
and the ratio
\be
R(\e)=\frac{\sum_{\a}f_{\a}(v+\e)}{\sum_{\a}\bar{f}_{\a}(v+\ve)}.
\ee
Substituting these results into Eq.~(\ref{current}) one finds 
$I_{L}+I_{R}=0$, as it should be.
Interestingly, for $V=0$ one recovers the 
thermal equilibrium distribution of the isolated impurity, i.e.,  
$P_{0}=1/\callZ$, $P_{1}=2e^{-\b v}/\callZ$ and 
$P_{2}=e^{-\b(2v+U)}/\callZ$, with $\callZ=1+2e^{-\b 
v}+e^{-\b(2v+U)}$ the partition function.

We next consider the (more general) time-dependent case and find an 
explicit solution for $N(t)$ and $I_{\a}(t)$. We write the 
probability vector $\overrightarrow{P}=(P_{0},P_{1},P_{2})^{T}$ as 
proposed in Ref.~\cite{dittmann2017non}
\be
\overrightarrow{P}=\left(
\begin{array}{c}
    1-N \\ N \\ 0
\end{array}
\right)+p\left(
\begin{array}{c}
    1 \\ -2 \\ 1
\end{array}
\right).
\label{paramP}
\ee
This parametrization satisfies Eqs.~(\ref{norm}) and (\ref{dens}) 
by construction. Using Eq.~(\ref{paramP}) to rewrite the RE one 
finds
\be
\frac{1}{\g}\frac{d}{dt}
\overrightarrow{P}=
\left(
\begin{array}{c}
    (N-1)F(v)+N\bar{F}(v)/2 \\
    (1-N)F(v)-N[F(v+U)-\bar{F}(v)]/2 \\ 
    NF(v+U)/2
\end{array}
\right),
\label{rev2}
\ee
where
\be
F(\e)=\sum_{\a}f_{\a}(\e)\quad,\quad
\bar{F}(\e)=\sum_{\a}\bar{f}_{\a}(\e).
\ee
Notice that, as in the case of the single reservoir considered in 
Ref.~\cite{dittmann2017non}, the right hand side is independent of $p$.

To obtain the impurity occupation we take the inner product of
Eq.~(\ref{rev2}) with the vector $(0,1,2)$, see Eq.~(\ref{dens}), and find
\be
\frac{\dot{N}}{\gamma}=-N+\left(1-\frac{N}{2}\right)F(v)+\frac{N}{2}F(v+U).
\label{tddens}
\ee
Equation~(\ref{tddens}) gives the correct steady-state occupation for $\dot{N}=0$.
For the current we consider the combination
\be
I=\frac{1}{2}(I_{L}-I_{R}) \;.
\ee
From knowledge of $\dot{N}$ and $I$ one can always extract $I_{L}$ and $I_{R}$
using the continuity equation $I_{L}+I_{R}=\dot{N}$. Taking into account 
Eq.~(\ref{current}) and the parametrization of Eq.~(\ref{paramP}) it 
is straightforward to find
\be
\frac{I}{\g}=\left(1-\frac{N}{2}\right)\D f(v)+\frac{N}{2}\D f(v+U),
\label{tdcurr}
\ee
with 
\be
\D f(\e)=f_{L}(\e)-f_{R}(\e).
\ee

With the occupation and current as functions of gate and bias we can 
invert the map for finite $U$ and for vanishing $U$. Denoting 
by $(v,V)$ the interacting (finite $U$) inverse map and by 
$(v_{s},V_{s})$ the noninteracting ($U=0$) inverse map, 
the  TD i-DFT potentials are given by 
\begin{subequations}
\bea
v_{\rm Hxc}[n,I]&\equiv&v_{s}[n,I]-v[n,I].
\\
V_{\rm xc}[n,I]&\equiv&V_{s}[n,I]-V[n,I].
\eea
\label{xcpot}
\end{subequations}  
The key observation to simplify the inversion problem is
that Eqs.~(\ref{tddens}, \ref{tdcurr}) can 
be written as 
\begin{subequations}
\bea
\frac{\dot{N}}{\gamma}+N=\int [f(\w-V/2)+f(\w+V/2)]A_{U}(\w-v)\quad
\\
I=\frac{\g}{2}\int 
[f(\w-V/2)-f(\w+V/2)]A_{U}(\w-v),\quad\quad
\eea
\label{AMmap}
\end{subequations}
with the spectral function
\be
A_{U}(\w)=2\p \left[\frac{N}{2}\d(\w-U)+(1-\frac{N}{2})\d(\w)\right].
\label{spectralfunction}
\ee
Therefore,
in terms of the variables 
$w_{\pm}=v\pm V/2$ the problem is separable since Eqs.~(\ref{AMmap})
read
\be
N+(\dot{N}\mp 2I)/\g=2\int f(\w)A_{U}(\w-w_{\pm})\equiv
\mathfrak{n}_{U}(w_{\pm}),
\label{sepvar}
\ee
which can be solved, e.g., by the bisection method. 

For $\dot{N}=0$, Eq.~(\ref{sepvar}) is the same equation used in
 Ref.~\cite{StefanucciKurth:15} to obtain the steady-state xc 
potentials of i-DFT. In that work we also took into account 
that the spectral peaks are broadened  due to the coupling to the 
leads and replaced the delta functions in 
Eq.~(\ref{spectralfunction}) with normalized Lorentzians of width 
$\gamma$. It was then shown that  an accurate 
parametrization
for the difference of the noninteracting and interacting 
inverse maps is
\bea
v_{\rm Hxc}\pm \frac{V_{\rm xc}}{2}
&=&
\mathfrak{n}^{-1}_{U=0}(N\mp 2I/\g)-\mathfrak{n}^{-1}_{U}(N\mp 2I/\g)
\nn\\
&\approx&\frac{U}{2}+\frac{U}{\p}{\rm atan}\left(\frac{N\mp I/\g 
-1}{W}\right),
\eea
with $W=0.16 \g/U$. Writing down the (H)xc potentials explicitly, we have
\begin{subequations}
\bea
v_{\rm Hxc}^{\rm ad1}[N,I]&=&\frac{U}{2}
\nn\\
&+&\frac{U}{2}\sum_{s=\pm}\frac{1}{\p}\,
{\rm 
atan}\!\left(\!\frac{N + sI/\g-1}{\lambda W}\right)
\quad\quad
\label{xcgatepar_nomem}
\\
V_{\rm xc}^{\rm ad1}[N,I]&=&-U\sum_{s=\pm}\frac{s}{\p}\,
{\rm atan}\left(\!\frac{N + sI/\g-1}{\lambda W}\right)
\label{xcbiaspar_nomem}
\eea
\label{xcpar_nomem}
\end{subequations}
where the superscript ``ad'' indicates that the functional will be used in
an adiabatic sense and we also have introduced, for later use, the
parameter $\lambda$ which here is set to unity, $\lambda=1$. 

As the mathematical structure of 
$\mathfrak{n}_{U}$ does not change for $\dot{N}\neq 0$, we can 
obtain a parametrization for the TD i-DFT potentials with memory by simply
replacing $I \to I\mp  \dot{N}/2$
\begin{subequations}
  \bea
  \lefteqn{
v_{\rm Hxc}^{\rm mem1}[N,I]=\frac{U}{2}}
\nn\\
&&+\frac{U}{2}\sum_{s=\pm}\frac{1}{\p}\,
{\rm 
atan}\!\left(\!\frac{N+(\frac{\dot{N}}{2}+sI)/\g-1}{\lambda W}\right)
\quad\quad
\label{xcgatepar}
\eea
\bea
\lefteqn{
  V_{\rm xc}^{\rm mem1}[N,I]=}
\nn\\
&&-U\sum_{s=\pm}\frac{s}{\p}\,
{\rm atan}\left(\!\frac{N+(\frac{\dot{N}}{2}+sI)/\g-1}{\lambda W}\right)
.\quad\quad
\label{xcbiaspar}
\eea
\label{xcpar}
\end{subequations}
Notice that, as anticipated,  the dependence on history occurs 
through the first time-derivative of the impurity occupation.
We further observe that along the ``physical'' solutions the 
continuity equation implies that $I_{L}=I+\dot{N}/2$ and 
$I_{R}=-I+\dot{N}/2$. Therefore, along the physical solutions we can 
rewrite the TD i-DFT potential as
\begin{subequations}
\bea
v_{\rm Hxc}^{\rm mem1}[N,I]&=&\frac{U}{2}+\frac{U}{2\p}
{\rm atan}\!\left(\!\frac{N+I_{L}/\g-1}{\lambda W}\right)
\nn\\
&+&\frac{U}{2\p}
{\rm atan}\!\left(\!\frac{N+I_{R}/\g-1}{\lambda W}\right)
\quad\quad
\label{xcgatepar2}
\\
V_{\rm xc}^{\rm mem1}[N,I]&=&\frac{U}{\p}\,
{\rm atan}\left(\!\frac{N+I_{R}/\g-1}{\lambda W}\right)
\nn\\
&-&
\frac{U}{\p}\,
{\rm atan}\left(\!\frac{N+I_{L}/\g-1}{\lambda W}\right)
.\quad\quad
\label{xcbiaspar2}
\eea
\label{xcpar2}
\end{subequations}
This is the form used for the time-dependent simulations of the next 
Section.

\section{Results}

We have performed time-dependent transport simulations for the Anderson model
using a modified version of the algorithm of Ref.~\cite{ksarg.2005} which
allows to take into account not only an Hxc gate potential
on the dot but also a (time-dependent) xc contribution to the bias both at 
zero and at finite temperatures. In our
TD simulations we use one-dimensional tight binding leads which means that our 
leads have a semi-circular density of states of finite bandwidth while 
the functionals described in the previous section were derived for the 
WBL case. However, by proper choice of parameters one can ensure
that the WBL is (approximately) achieved: at equilibrium, the leads 
are taken to be at half-filling, the coupling between the impurity and the 
leads is taken sufficiently small, and all relevant energies (on-site gate 
potential, electron-elec\-tron interaction) are well within the bandwidth 
of the lead bands. 
We are interested in the effects of the memory term (i.e. $\dot{N}$) in the 
functionals. We study two distinct situations, a bias quench and a quench of 
the gate potential. 

\begin{figure}[t]
  \includegraphics[width=0.47\textwidth]{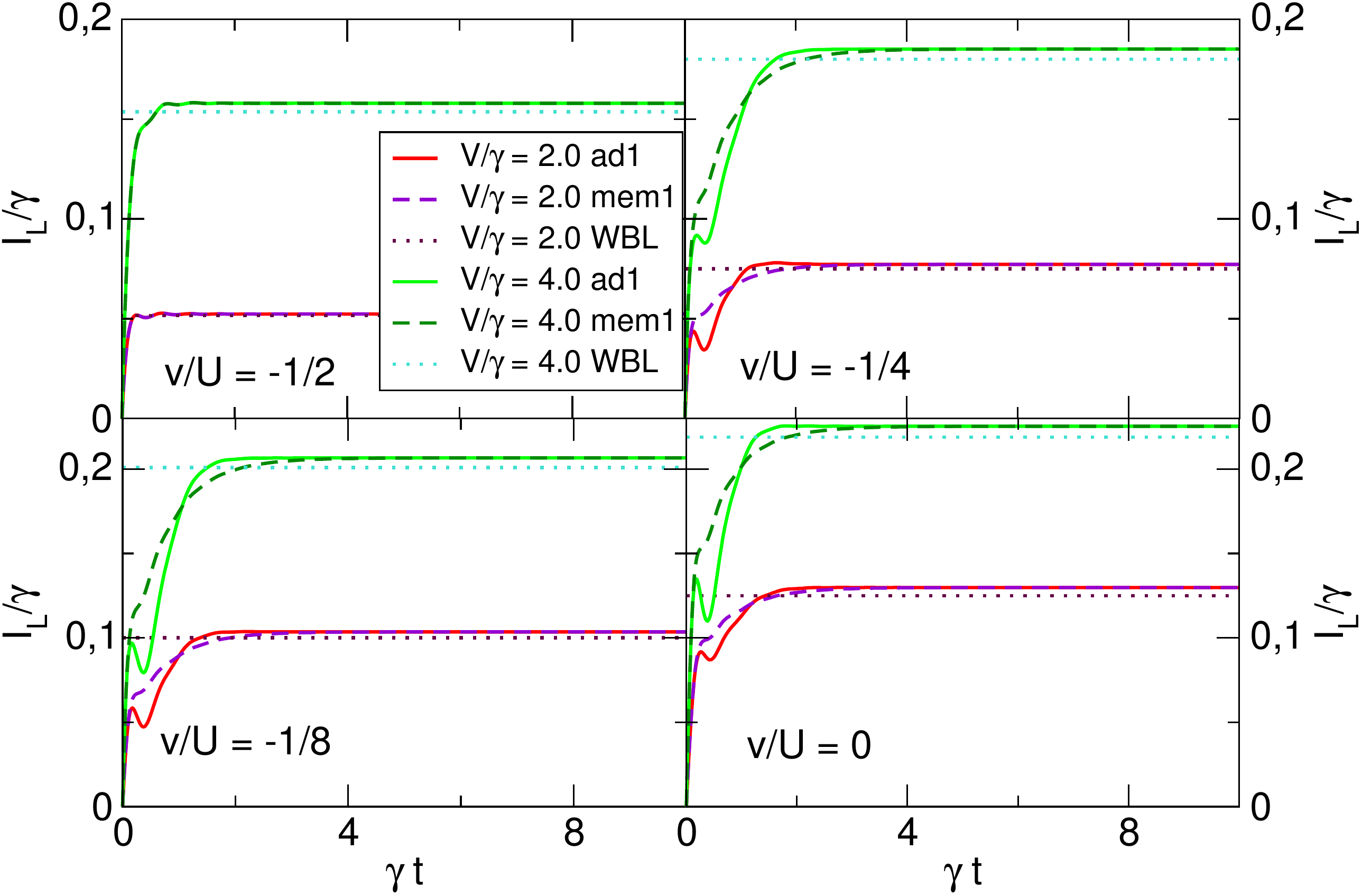}
  \caption{Time-dependent currents from left lead to the Anderson
    impurity after a bias quench at time $t=0$. Results are shown both for
    the adiabatic (ad1) and memory functionals (mem1) of
    Eqs.~(\ref{xcpar_nomem}) and (\ref{xcpar2}), respectively. Different
    panels correspond to different values of the gate potential. Steady
    currents in the wide band limit (WBL) are also shown for comparison.
    Parameters: $U/\g=3$ for the interaction and $T/\g=1.0$ for the
    temperature. }
\label{comp_SK15_mem}
\end{figure}

In Fig.~\ref{comp_SK15_mem} we show the TD current $I_L$ after a bias quench, 
i.e. after switching on a bias of the form $V(t) = V \theta(t)$ where 
$\theta(t)$ is the Heaviside step function. For times $t<0$ the system is 
in equilibrium. The bias is applied symmetrically 
in left and right leads, $V_L(t)=-V_R(t) = V(t)/2$. The different panels 
of Fig.~\ref{comp_SK15_mem} show the currents obtained from the functional 
without and with memory, Eqs.~(\ref{xcpar_nomem}) and (\ref{xcpar2}),
respectively, for different values of the (static) gate potential $v$ at 
fixed temperature $T/\g=1.0$. We
also show the steady state currents (dashed lines) obtained from the
functional of Eq.~(\ref{xcpar_nomem}) in the WBL. As it should be, in the
long-time limit both functionals lead to the same steady currents which agree
well with those of the WBL, confirming that we have chosen our parameters
properly. 

As for the TD currents, we see that at the particle-hole symmetric point 
$v=-U/2$ (upper left panel), both functionals by symmetry 
give exactly the same TD current. This is easily understood by the fact 
that in this particular case the density is constant, $N=1$, at all times 
and thus $\dot{N}=0$ and the two functionals become completely equivalent. 
For other values of the gate, both functionals lead to slightly different 
TD currents where the transient current oscillations in $I_L$ tend to be
more pronounced for the functional without memory.

Both the functionals of Eqs.~(\ref{xcpar_nomem}) and (\ref{xcpar2}) are derived
from the rate equations and thus contain Coulomb blockade but no Kondo physics.
In Ref.~\cite{KurthStefanucci:16} we have designed a steady-state functional
which also incorporates Kondo physics. At zero temperature this functional
can be written as
\begin{subequations}
\bea
v_{\rm Hxc}^{\rm ad2}[N,I] &=& \left( 1-a[I] \right) v_{\rm Hxc}^{\rm ad1}[N,I]
+ a[I] v_{\rm Hxc}^{(0)}[N]
\quad\quad
\label{xcgate_kondo_nomem}
\\
V_{\rm xc}^{\rm ad2}[N,I] &=& \left( 1-a[I] \right) V_{\rm xc}^{\rm ad1}[N,I]
\label{xcbias_kondo_nomem}
\eea
\label{xc_kondo_nomem}
\end{subequations}
where $a[I]$ is defined as
\be
a[I] = 1 - \left[ \frac{2}{\pi} \arctan\left( \frac{I}{\g W}\right) \right]^2
\ee
and $v_{\rm Hxc}^{(0)}[N]$ is the accurate parametrization of the SIAM Hxc
potential at zero temperature given in Ref.~\cite{blbs.2012}. Furthermore,
in both $v_{\rm Hxc}^{\rm ad1}$ and $V_{\rm xc}^{\rm ad1}$, the value of the
parameter $\lambda$ is set to $\lambda=2$. The above functional (and its
extension to finite temperature) has been
used in Ref.~\cite{stefanucci2017ac} to study the influence of an AC bias
on the DC conductance of the SIAM. Since Eqs.~(\ref{xc_kondo_nomem})
derive from a steady-state functional, they do not contain any memory.
In order to include memory, we propose the analogous form
\begin{subequations}
\bea
v_{\rm Hxc}^{\rm mem2}[N,I] &=& \left( 1-a[I] \right) v_{\rm Hxc}^{\rm mem1}[N,I]
+ a[I] v_{\rm Hxc}^{(0)}[N]
\quad\quad
\label{xcgate_kondo_mem}
\\
V_{\rm xc}^{\rm mem2}[N,I] &=& \left( 1-a[I] \right) V_{\rm xc}^{\rm mem1}[N,I]
\label{xcbias_kondo_mem}
\eea
\label{xc_kondo_mem}
\end{subequations}
where, again, we choose $\lambda=2$ in both $v_{\rm Hxc}^{\rm mem1}$ and
$V_{\rm xc}^{\rm mem1}$. 

\begin{figure}[t]
  \includegraphics[width=0.47\textwidth]{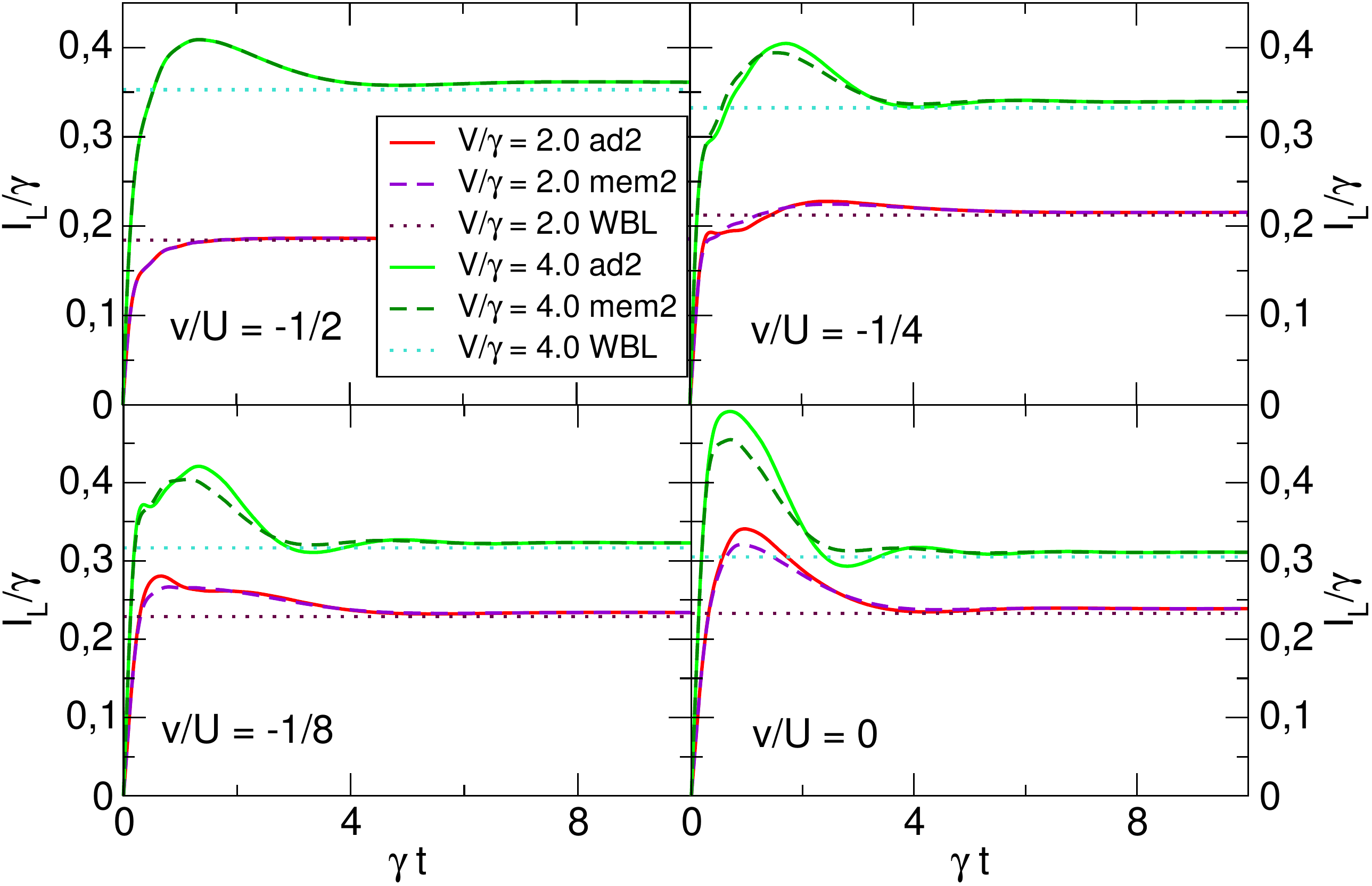}
  \caption{Same as Fig.~\ref{comp_SK15_mem} but now at $T/\g=0$ and using
    the functionals of Eqs.~(\ref{xc_kondo_nomem}) and (\ref{xc_kondo_mem})
    valid in the Kondo regime. Other parameters as before.}
\label{comp_KS16_mem}
\end{figure}

In Fig.~\ref{comp_KS16_mem} we show the TD currents obtained with the
functionals (\ref{xc_kondo_nomem}) and (\ref{xc_kondo_mem}) as a result of a
bias quench. Again, the TD currents for both functionals correctly approach the
steady currents of the WBL in the long-time limit. Also, at the ph symmetric
point both functional again give the same time evolved currents, exactly for
the same reason as before: in this case $N=1$ and thus $\dot{N}=0$ at all
times, rendering the two functionals equivalent. Away from ph symmetry,
the TD currents for both functionals are qualitatively quite similar,
the memory functional leading to slightly less pronounced transient
oscillations in $I_L$ as also was the case for the functionals of
Eqs.~(\ref{xcpar_nomem}) and (\ref{xcpar2}).

\begin{figure}[t]
  \includegraphics[width=0.47\textwidth]{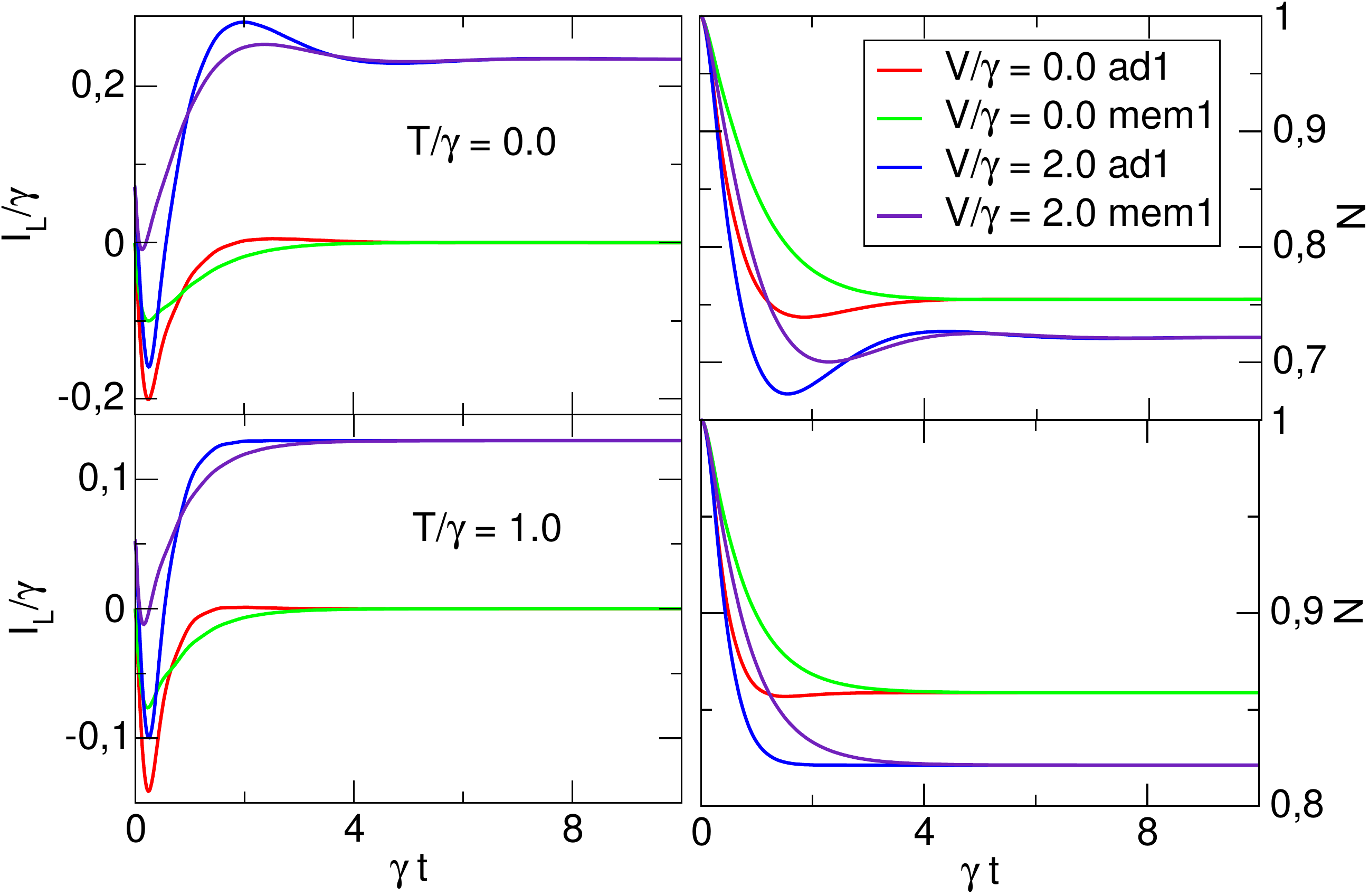}
  \caption{Comparison of TD currents (left) and TD densities (right) 
    for the SIAM with $U/\g=3$ obtained with the functionals of
    Eqs.~(\ref{xcpar_nomem}) and (\ref{xcpar2}) after a sudden quench (at
    time $t=0$) of the gate potential. For times $t<0$, the gate is fixed at
    the ph symmetric point $v=-U/2$, at $t=0$ it is suddenly switched to $v=0$.
    For the biased case ($V/\g=2.0$), the bias was switched on at some time
    (negative) $t_0$ and the system was then propagated until a steady
    state was reached before, at $t=0$, the gate potential was quenched. 
    The calculations are done for two different temperatures: $T/\g=0.0$ 
    (upper panels) and $T/\g=1.0$ (lower panels). 
  }
\label{comp_quench_gate_SK15}
\end{figure}

In Fig.~\ref{comp_quench_gate_SK15} we show the time evolution of the
current and the density on the dot after a quench of the {\em gate} potential
at time $t=0$ using the functionals of Eqs.~(\ref{xcpar_nomem}) and
(\ref{xcpar2}) for two different temperatures. For times $t<0$, the gate 
potential is kept constant at the ph symmetric point, $v=-U/2$. In the 
unbiased case ($V/\g=0$), the time evolution starts from equilibrium when, 
at $t=0$, the gate potential is suddenly quenched to $v=0$ (and afterwards 
kept constant). For the biased case ($V/\g=2$), we also start from 
equilibrium, but at some distant (negative) time $t_0$ we suddenly switch 
on a bias and evolve the system towards its steady state. Thus, when the 
gate is quenched at $t=0$, the propagation starts from this non-equilibrium 
state. Strictly speaking, the functionals (\ref{xcpar_nomem}) and 
(\ref{xcpar2}) are valid only for 
temperatures $T\geq \g$. However, we have still performed some calculations 
for $T=0$ (upper panels) in order to show that the transients at 
low temperature are much more pronounced than at finite temperature 
($T/\g=1.0$, lower panels). As in the previous cases and for both 
temperatures, the functional with memory leads to smaller transient 
oscillations both in the current and the density. 

\section{Conclusions}

The central ingredients for an i-DFT description of steady-state transport
through a nanoscale region connected to two leads are the Hxc gate and xc
bias potentials which are functionals both of the density in that region as
well as the steady current through it
\cite{StefanucciKurth:15,KurthStefanucciJPCreview}. For the SIAM, the first
suggested i-DFT functional \cite{StefanucciKurth:15} is  valid in the
CB regime. However, later a functional has been proposed
\cite{KurthStefanucci:16} which yields accurate results for a wide range
of parameters not only in the CB but also in the Kondo regime.

These i-DFT steady-state functionals have been generalized in an adiabatic
sense to explicitly time-dependent transport \cite{stefanucci2017ac}. 
In the present work we generalize a recently proposed idea
\cite{dittmann2017non} to include memory effects in the xc gate potential to
the two-lead transport setup, i.e., both Hxc gate and xc bias include
memory. The inclusion of memory is achieved via reverse engineering of the
time-dependent rate equations \cite{rate-paper1} and thus at first only
valid in the CB regime. Since the i-DFT functional for the CB regime
serves as an essential ingredient in the construction of the accurate
SIAM functional for both Kondo and CB regime, the memory term can also
be included in this functional. We have shown with explicit examples of
TD transport simulations for the SIAM, that the adiabatic and
memory functional lead to different transient evolutions in density and
current but, by construction, give the same steady state. 

\section*{Acknowledgements}
S.K. acknowledges funding by a grant of the "Ministerio de Economia y
Competividad (MINECO)" (FIS2016-79464-P) and by the
``Grupos Consolidados UPV/EHU del Gobierno Vasco'' (IT578-13). 
G.S. acknowledges funding by MIUR FIRB Grant No. RBFR12SW0J, EC funding
through the RISE Co-ExAN (GA644076), and funding through the INFN-Nemesys
project.

\end{document}